\documentclass[aps,pra,twocolumn,10pt,nofootinbib]{revtex4-1}
\usepackage{epsfig}
\usepackage{dcolumn}
\usepackage{natbib}
\usepackage[colorlinks]{hyperref}
\usepackage[utf8]{inputenc}
\usepackage[english]{babel}
\usepackage{graphicx}
\usepackage{xcolor}
\usepackage{bm}
\usepackage{amssymb}
\usepackage[intlimits]{amsmath}
\usepackage{amsmath}
\usepackage{braket}
\usepackage[para,online,flushleft]{threeparttable}

\begin{document}

\title{Time- and parity-violating effects of nuclear Schiff moment in molecules and solids}
\author{V. V. Flambaum$^{1,2}$}
\author{V. A. Dzuba$^1$}
\author{H. B. Tran Tan$^1$}
\affiliation{$^1$School of Physics, University of New South Wales,
Sydney 2052, Australia}
\affiliation{$^2$Helmholtz Institute Mainz, Johannes Gutenberg University, 55099 Mainz, Germany}

\begin{abstract}
We show that existing calculations of the interaction between nuclear Schiff moments and electrons in molecules use an inaccurate operator which gives rise to significant errors. By comparing the matrix elements of the accurate and imprecise Schiff moment operators, we calculated the correction factor as a function of the nuclear charge Z and presented corrected results for the T,P-violating interaction of the nuclear spin with the molecular axis in the TlF, RaO, PbO, TlCN, ThO, AcF molecules and in the ferroelectric solid PbTiO$_3$.


\end{abstract}
\date{\today}
\maketitle

\section{Introduction}
The search for new interactions which violate both the time-reversal (T) and parity (P) invariances is of fundamental importance for the study of Physics beyond the Standard Model. It is well known that, among other phenomena, P,T-odd nuclear interactions give rise to the nuclear Schiff moment\ \cite{Sandars1967,Hinds1980,SFK,FLAMBAUM1986}, which may interact with electrons and cause measurable shifts in atomic and molecular spectra. 

Currently, all calculations of the interaction of nuclear Schiff moments with electrons in molecules use an operator which is only approximately correct\ \cite{Hinds1980,Parpia1997,Quiney1998,Petrov2002,Flambaum2002,Kudashov2013,Skripnikov2016,atoms7030062}. Although a more accurate form for this operator do exist\ \citep{FlambaumGinges2002}, its use has not been widely adopted. In this paper, we demonstrate that the imprecise operator may lead to significant inaccuracy of the results and suggest a simple way to amend the situation.

The rest of the paper will be organized as follows. In Sect.\ \ref{forms}, we give a brief review of the various forms of interaction and the problems associated with them. We will also introduce scaling factors, which are the ratios between the matrix elements of the correct and imprecise operators. These factors may be used to correct the published results obtained from the imprecise operator. In Sect.\ \ref{Ana} and\ \ref{Num}, we present the analytical and numerical results for these scaling factors. Sect.\ \ref{Conclusion} contains a short summary of our findings.
\section{The different forms of interaction}\label{forms}

For a pointlike nucleus, the electro-static potential produced by the Schiff moment is presented in the form\ \cite{Sandars1967,SFK} \footnote{Ref.\ \cite{Sandars1967} considered the effects of proton electric dipole moment whereas Ref.\ \cite{SFK} consider the effects of P,T-odd nuclear forces which give a bigger contribution to the Schiff moment.}
\begin{equation}\label{center}
\varphi_{0}\left({\bf R}\right)=4\pi{\bf S}\cdot\nabla\delta\left({\bf R}\right)\,,
\end{equation}
where $\bf R$ is the electron position vector from the center of the nucleus, $\delta\left({\bf R}\right)$ is the delta function and $\bf S$ is the nuclear Schiff moment.

Let us consider the matrix element of $-e\varphi_0$ between the electronic \textit{s} and \textit{p} states, which reads
\begin{equation}\label{M_center}
\begin{aligned}
M_{0}^{sp}=4\pi{\bf S}\cdot\left(\nabla\psi_s^{\dagger}\psi_p\right)_{R\rightarrow 0}\,,
\end{aligned}
\end{equation}
where $\psi_{s,p}$ are the $s$ and $p$ wavefunction for an electron in the potential of a point-like nucleus.

We note that although the quantity $\left(\nabla\psi_s^{\dagger}\psi_p\right)_{R\rightarrow 0}$ tends toward a constant value for a non-relativistic electron, it becomes infinite for a relativistic electron. This issue is resolved by considering a finite-size instead of a point-like nucleus. One solution is to simply cut off the point-like-nucleus relativistic electron wavefunction at the surface of the nucleus, obtaining  
\begin{equation}\label{M_surface}
\begin{aligned}
M_{1}^{sp}=4\pi{\bf S}\cdot\left(\nabla\psi_s^{\dagger}\psi_p\right)_{R\rightarrow R_N}\,,
\end{aligned}
\end{equation}
where $R_N$ is the radius of the nucleus. However, since the variation of $\nabla\psi_s^{\dagger}\psi_p$ inside the nucleus is of the order of $Z^2\alpha^2$ where $Z$ is the nuclear charge and $\alpha$ is the fine-structure constant, formula\ \eqref{M_surface} does not give reliable results in the case of a heavy nucleus. 

Another approach is to use, for the wavefunctions $\psi_{s,p}$, the correct solutions to the Dirac equation with the potential from a finite-size nucleus. Although in this case, the quantity $\left(\nabla\psi_s^{\dagger}\psi_p\right)_{R\rightarrow 0}$ is finite, using Eq.\ \eqref{M_center} to calculate the matrix element is logically inconsistent (since Eq.\ \eqref{M_center} corresponds to the Schiff moment being placed at the center of the nucleus) and may give imprecise results. Unfortunately, this approach in used in all molecular calculations of the effects of nuclear Schiff moments.

A more accurate formula for the potential produced by the nuclear Schiff moment was derived in Ref.~\cite{FlambaumGinges2002}
\begin{equation}\label{e:S1}
\varphi_2(\mathbf{R}) = \frac{3\mathbf{S\cdot R}}{B}n(R)\,,
\end{equation}
where $B=\int n\left(R\right)R^4dR\approx R_N^5$ and $n\left(R\right)$ is nuclear charge density. This formula correspond to a constant electric field ${\bf E}=-\nabla\varphi_2$ produced by the Schiff moment at the center of the nucleus and zero field outside the nucleus. Here, we see that the actual interaction with the Schiff moment potential vanishes at the center of the nucleus which explains why placing the Schiff moment at the center of the nucleus as in Eq.\ \eqref{M_center} is not reasonable. However, all molecular calculations used Eq.\ \eqref{M_center}.

The corresponding matrix element of $-e\varphi_2$ between $s$ and $p$ waves is given by
\begin{equation}\label{M_correct}
M^{sp}_2=\frac{3{\bf S}}{B}\cdot\int\psi^{\dagger}_s{\bf R}\psi_pn\left(R\right)d^3R\,,
\end{equation}
where $\psi_{s,p}$ is the electron wavefunction corresponding to a finite-size nucleus.

As we will show below, compared to the correct formula\ \eqref{M_correct}, Eqs.\ \eqref{M_center} and\ \eqref{M_surface} give rise to errors as large as 20\%, which may compromise the reliability of otherwise very accurate atomic and molecular calculations. In this paper, we present a solution to this problem. We provide the numerical values for the ratios of the matrix elements
\begin{equation}\label{ratio_define}
\begin{aligned}
r^{sp}_{\rm center}=\left|\frac{M^{sp}_0}{M^{sp}_2}\right|\,\,\,{\rm and}\,\,\,r^{sp}_{\rm surface}=\left|\frac{M^{sp}_1}{M^{sp}_2}\right|\,,
\end{aligned}
\end{equation}
which can be used to simply rescale any result involving the matrix elements\ \eqref{M_center} and\ \eqref{M_surface} to the correct value which involves the matrix element\ \eqref{M_correct}.

\section{Analytical results}\label{Ana}
The matrix element\ \eqref{M_center} may be written as\ \citep{FlambaumGinges2002}
\begin{equation}\label{appendix_center}
\begin{aligned}
M_{\rm center}^{sp}=3e{\bf S}\cdot\bra{s}{\bf n}\ket{p}\left.\frac{U_{sp}\left(R\right)}{R}\right|_{R\rightarrow 0}\,,
\end{aligned}
\end{equation}
where ${\bf n}={\bf R}/R$, $\bra{s}{\bf n}\ket{p}=\int\Omega_s^{\dagger}{\bf n}\Omega_p\sin\theta d\theta d\phi$ and $U_{sp}\left(R\right)=f_s\left(R\right)f_p\left(R\right)+g_s\left(R\right)g_p\left(R\right)$. Here, $f_{s,p}$ and $g_{s,p}$ are the upper and lower component of the electron wave function.

Similarly, the matrix element\ \eqref{M_surface} may be written as
\begin{equation}\label{appendix_surface}
\begin{aligned}
M_{\rm surface}^{sp}=3e{\bf S}\cdot\bra{s}{\bf n}\ket{p}\left.\frac{U_{sp}\left(R\right)}{R}\right|_{R\rightarrow R_N}\,,
\end{aligned}
\end{equation}
and the matrix element\ \eqref{M_correct} as
\begin{equation}\label{appendix_correct}
\begin{aligned}
M_{\rm body}^{sp}&=\frac{15\bf S}{R_N^5}\cdot\bra{s}{\bf n}\ket{p}\int U_{sp}nR^3dR\\
&\approx\frac{15\bf S}{R_N}\cdot\bra{s}{\bf n}\ket{p}\int_0^1 U_{sp}\left(x\right)x^3dx\,,
\end{aligned}
\end{equation}
where we have, for simplicity, assumed that the nucleus has a sharp surface, i.e., $n\left(R-R_N\right)=\theta\left(R_N-R\right)$ where $\theta$ is the Heaviside function. A more accurate numerical calculation will be performed in the next section. We found that the difference is insignificant.

Using Eqs.\ \eqref{appendix_center},\ \eqref{M_surface} and\ \eqref{M_correct}, we may write
\begin{subequations}
\begin{align}
r_{\text{center}}^{sp}&=\frac{{{\left. {{U}_{sp}}\left( R \right)/R \right|}_{R\to 0}}}{\frac{5}{{{R}_{N}}}\int\limits_{0}^{1}{{{U}_{sp}}\left( x \right){{x}^{3}}dx}}\,,\label{rcenter}\\ 
r_{\text{surface}}^{sp}&=\frac{{{\left. {{U}_{sp}}\left( R \right)/R \right|}_{R\to {{R}_{N}}}}}{\frac{5}{{{R}_{N}}}\int\limits_{0}^{1}{{{U}_{sp}}\left( x \right){{x}^{3}}dx}}\,.\label{rsurface}
\end{align}
\end{subequations}

The analytical form of the functions $U_{sp_{1/2}}$ and $U_{sp_{3/2}}$ for a finite nucleus may be taken from Ref.\ \cite{FlambaumGinges2002}. They read\footnote{Because we calculate ratios of the matrix elements, the normalization of these functions is not important. These functions are calculated to the fourth order of $Z\alpha$, which provides sufficient accuracy, as can be seen from numerical calculations described in the next section.}
\begin{subequations}
\begin{align}
{{U}_{s{{p}_{1/2}}}}\left(x\right)&\sim {{R}_{N}}x\left\{ 1-\frac{3}{5}{{\left( Z\alpha  \right)}^{2}}{{x}^{2}}\left[ 1-\frac{3}{14}{{x}^{2}}\right.\right.\nonumber\\ 
&\left.\left.+\frac{2}{135}{{x}^{4}} \right]+\frac{81}{560}{{\left( Z\alpha  \right)}^{4}}{{x}^{4}} \right\}\,,\\
{{U}_{s{{p}_{3/2}}}}\left( x \right)&\sim x\left\{ 1-\frac{9}{20}{{\left( Z\alpha  \right)}^{2}}{{x}^{2}}\left[ 1-\frac{69}{315}{{x}^{2}}\right.\right.\nonumber\\
&\left.\left.+\frac{1}{63}{{x}^{4}} \right]+\frac{243}{2800}{{\left( Z\alpha  \right)}^{4}}{{x}^{4}} \right\}\,,
\end{align}
\end{subequations}
which, upon insertion into Eqs.\ \eqref{rcenter} and\ \eqref{rsurface}, give
\begin{subequations}\label{ratiosp1/2}
\begin{align}
r_{\text{Center}}^{s{{p}_{1/2}}}&\approx \frac{1}{1-0.361{{\left( Z\alpha  \right)}^{2}}+0.08{{\left( Z\alpha  \right)}^{4}}}\,,\label{a}\\
r_{\text{Surface}}^{s{{p}_{1/2}}}&\approx \frac{1-0.480{{\left( Z\alpha  \right)}^{2}}+0.145{{\left( Z\alpha  \right)}^{4}}}{1-0.361{{\left( Z\alpha  \right)}^{2}}+0.08{{\left( Z\alpha  \right)}^{4}}}\,,\label{b}\\
r_{\text{Center}}^{s{{p}_{3/2}}}&\approx \frac{1}{1-0.270{{\left( Z\alpha  \right)}^{2}}+0.05{{\left( Z\alpha  \right)}^{4}}}\,,\label{c}\\
r_{\text{Surface}}^{s{{p}_{3/2}}}&\approx \frac{1-0.359{{\left( Z\alpha  \right)}^{2}}+0.09{{\left( Z\alpha  \right)}^{4}}}{1-0.270{{\left( Z\alpha  \right)}^{2}}+0.05{{\left( Z\alpha  \right)}^{4}}}\,.\label{d}
\end{align}
\end{subequations}

\begin{figure}[htb]
    \centering
    \includegraphics[scale=0.32]{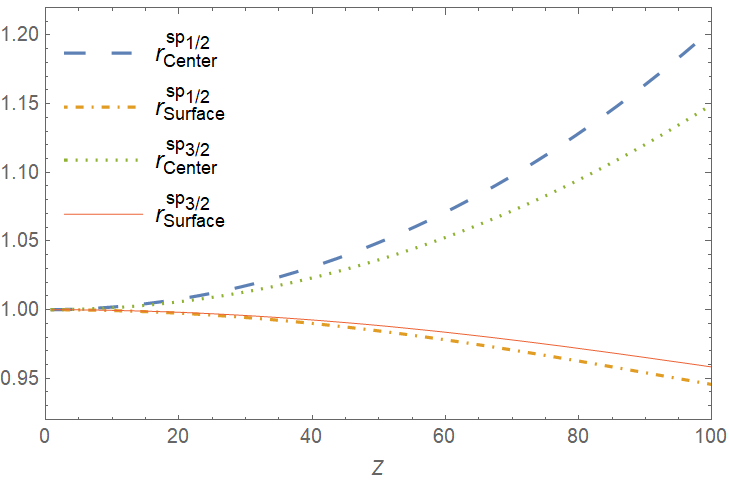}
    \caption{Ratios of the matrix elements of at-the-center\ \eqref{M_center}, on-the-surface\ \eqref{M_surface} and finite-size\ \eqref{M_correct} nucleus Schiff moment potentials as functions of the nuclear charge $Z$. The results are presented separately for $sp_{1/2}$ and $sp_{3/2}$ matrix elements.}
    \label{PlotRatios}
\end{figure}

The plots of these ratios as functions of the nuclear charge $Z$ are shown in Fig.\ \ref{PlotRatios}.

As can be observed from these plots, for small $Z$, the ratio $r^{sp}_{\rm center}$ and $r^{sp}_{\rm surface}$ tend to unity which indicates that formulae\ \eqref{M_center} and\ \eqref{M_surface} give reasonable estimates of the correct matrix elements. However, for large $Z$, the formula\ \eqref{M_center} presented an overestimate whereas the formula\ \eqref{M_surface} presents an underestimate of the result.

Also, the approximation\ \eqref{M_surface} which corresponds to placing the Schiff moment on the surface of the nucleus appears to have a better accuracy (error $\leq$ 5\%) than the approximation\ \eqref{M_center} which places the Schiff moment at the center of the nucleus (error may be as large as 20\%). The surface formula\ \eqref{M_surface} give a better accuracy since the accurate potential\ \eqref{e:S1} is zero at the center of the nucleus and has a maximum closer to the nuclear surface.
 
\section{Numerical calculation}\label{Num}
In this section, we provide the numerical results of our calculations for the ratio $r^{sp}_{\rm center}$ in several atoms and molecules of interest. Since the surface formula\ \eqref{M_surface} is not commonly used in molecular calculations, and since the inaccuracy associated with it is not severe (less than 5\%), we do not evaluate the correction factor $r^{sp}_{\rm surface}$. We use the obtained $r^{sp}_{\rm center}$ ratios to adjust the existing results of the interaction constant $W_S$ for the effective T,P-odd interaction in molecules defined by
\begin{equation}
H^{T,P}_{\rm eff}=W_S{\bf S}\cdot\hat{\bf n}
\end{equation}
where $\hat{\bf n}$ is the unit vector along the molecular axis\footnote{Some older molecular calculations presented the constant $X=W_S/6$.}.



For numerical calculations we use a standard Fermi distribution for $n\left(R\right)$\ \cite{Hofstadter1953,Hahn1956}
\begin{equation}
n\left(R\right)=\frac{a}{1+\exp\frac{R-R_N}{\delta}}.
\label{e:rho}
\end{equation}
where $a$ is a normalization constant chosen fixed by the condition $\int n dV = Z$ and $\delta$ is the nuclear skin thickness, $\delta\approx \delta_0/4\ln3$, $\delta_0=$2.3~fm.

Note that a good compromise between\ \eqref{center} and\ \eqref{e:S1} can be achieved if the delta function in\ \eqref{center} is replaced by Fermi distribution\ \eqref{e:rho}. The difference in the results in this case is reduced to just a few percent.

Numerical calculations of the correction factors may be done  at the relativistic Hartree-Fock-Dirac (HFD) level. At this level the matrix elements of the Schiff moment operator for higher waves (beyond s and p) are practically zero. All many-body corrections 
may be formally expressed as (infinite) sums of the HFD matrix elements.

To check the role of the many-body effects we use the so-called random-phase approximation (RPA).
The RPA equations present a linear response of the Hartree-Fock-Dirac atomic states to a perturbation by an external field. They can be written in a form
\begin{equation}\label{eq:RPA}
(H_0 -\epsilon_c)\delta \psi_c = -(F+\delta V^F)\psi_c.
\end{equation}
Here $H_0$ is the relativistic Hartree-Fock-Dirac Hamiltonian, $\psi_c$ is the HFD electron state in the core, $\delta \psi_c$ is the correction to the HFD state in the core induced by the external field $F$ given by (\ref{center}) or (\ref{e:S1}), $\delta V^F$ is the correction to the self-consistent HFD
potential due to the corrections to all core states. Index $c$ numerates states in the core. 
RPA equations (\ref{eq:RPA}) are solved self-consistently for all states in the core. 
It turns our that in spite of the fact that the RPA corrections increase matrix elements significantly, their ratio for the operators (\ref{center}) and (\ref{e:S1}) in Hartree-Fock-Dirac and RPA approximations is practically the same.

The results of atomic calculations for the ratios $r^{sp_{1/2}}_{\rm center}$ and $r^{sp_{3/2}}_{\rm center}$ are presented in Table~\ref{t:mol}. For comparison, the values obtained from Eqs.\ \eqref{a} and\ \eqref{c} are also presented. It is observed that the analytical formulae\ \eqref{ratiosp1/2} give results which are in good agreement with numerical calculations. 

The published and corrected values for the interaction constant $W_S$ of the effective T,P-odd interaction between the molecular electrons and the nuclear Schiff moments for several molecules are presented in Table~\ref{t:medm}. Note that we have taken the scaling factor as an average of $r^{sp_{1/2}}_{\rm center}$ and $r^{sp_{3/2}}_{\rm center}$, $\bar{r}^{sp}=\left(r^{sp_{1/2}}+r^{sp_{3/2}}\right)/2$. Indeed, the expansion of molecular orbitals over atomic orbitals centered at a heavy atom contains both $p_{1/2}$ and $p_{3/2}$ components. Because the factors $r^{sp_{1/2}}$ and $r^{sp_{3/2}}$ are close in value, the exact values of the weighting coefficients before the two components are not important. 

\begin{table}
\caption{\label{t:mol}Numerical values (n.) of the ratios of the $sp$ matrix elements with operators\ \eqref{center} and\ \eqref{e:S1}; $r^{sp_{1/2}}_{\rm center}=\left| \bra{s_{1/2}}\phi_0\ket{p_{1/2}}/\bra{s_{1/2}}\phi_2\ket{p_{1/2}}\right|$ and $r^{sp_{3/2}}_{\rm center}=\left| \bra{s_{1/2}}\phi_0\ket{p_{3/2}}/\bra{s_{1/2}}\phi_2\ket{p_{3/2}}\right|$. For comparison, the values obtained from the analytical (a.) formulae\ \eqref{a} and\ \eqref{c} are also presented.}
\begin{ruledtabular}
\begin{tabular}{ccdddd}
\multicolumn{1}{c}{$Z$}&
\multicolumn{1}{c}{Atom}&
\multicolumn{1}{c}{$r^{sp_{1/2}}_{\rm center}$ (n.)}&
\multicolumn{1}{c}{$r^{sp_{1/2}}_{\rm center}$ (a.)}&
\multicolumn{1}{c}{$r^{sp_{3/2}}_{\rm center}$ (n.)}&
\multicolumn{1}{c}{$r^{sp_{3/2}}_{\rm center}$ (a.)}\\
\hline
70 & Yb  & 1.11 & 1.10 & 1.09 & 1.07 \\
81 & Tl  & 1.15 & 1.13 & 1.11 & 1.10 \\
80 & Pb  & 1.15 & 1.13 & 1.12 & 1.09 \\
88 & Ra  & 1.18 & 1.16 & 1.13 & 1.12 \\
89 & Ac  & 1.18 & 1.16 & 1.14 & 1.12 \\
90 & Th  & 1.19 & 1.16 & 1.14 & 1.12 \\
\end{tabular}
\end{ruledtabular}
\end{table}

\begin{table}
\caption{\label{t:medm}The interaction constants $W_S=6X$ between the nuclear Schiff moment and the molecular axis. Published values (p.) were obtained with the\ \eqref{center} operator. Corrected values (c.) are obtained by dividing the published results by the factor $\bar{r}^{sp}=\left(r_{\rm center}^{sp_{1/2}}+r_{\rm center}^{sp_{3/2}}\right)/2$. All values of $W_S$ are given in a.u.}
\begin{ruledtabular}
\begin{tabular}{ccccc}
\multicolumn{1}{c}{Molecule}&
\multicolumn{1}{c}{$W_S$ (p.)}&
\multicolumn{1}{c}{Ref.}&
\multicolumn{1}{c}{$\bar{r}^{sp}$}&
\multicolumn{1}{c}{$W_S$ (c.)}\\
\hline
TlF       & 14000 & \cite{Hinds1980}                                      & 1.13 & 12400\\
TlF       & 46458 & \cite{Parpia1997}                                     & 1.13 & 41113\\
TlF       & 52482 & \cite{Quiney1998}                                     & 1.13 & 46444\\
TlF       & 45810 & \cite{Petrov2002}                                     & 1.13 & 40539\\
RaO       & 45192 & \cite{Flambaum2002,Kudashov2013}                      & 1.16 & 39127\\
PbTiO$_3$ & 30270 & \cite{Skripnikov2016}                                 & 1.14 & 26670\\
PbO       & 44400 & \cite{Skripnikov2016}                                 & 1.14 & 39119\\
TlCN      & 7150  & \cite{atoms7030062}                                   & 1.13 &  6327\\
ThO       & 45000 & \cite{FlambaumDzuba2019}                              & 1.17 & 39000\\
AcF       & 160000& \cite{FlambaumDzuba2019}                              & 1.16 & 140000\\
\end{tabular}
\end{ruledtabular}
\end{table}

\section{Conclusion}\label{Conclusion}
We showed in this paper how the use of an imprecise form for the interaction between the nuclear Schiff moment and the atomic and molecular electrons overestimates the results of molecular calculations up to 20\%. We provided the analytical formulae tested by the numerical calculations for the ratio between the electronic matrix elements using the correct and imprecise operators. We presented the corrected results for all existing molecular calculations. Our scaling factors may be used to obtain correct results in future molecular calculations using existing computer codes.

\section*{Acknowledgements}
This work was supported by the Australian Research Council and the Gutenberg Fellowship.

\bibliographystyle{apsrev}
\bibliography{bib}
\end{document}